\documentclass[ecta, final]{econsocart}
\RequirePackage[colorlinks,citecolor=blue,linkcolor=blue,urlcolor=blue,pagebackref]{hyperref}

\startlocaldefs

\theoremstyle{plain}
\newtheorem{corollary}{Corollary}
\newtheorem{axiom}{Axiom}

\theoremstyle{definition}
\newtheorem{definition}{Definition}
\newtheorem{proposition}{Proposition}
\newtheorem{remark}{Remark}
\endlocaldefs

\begin{document}

\begin{frontmatter}

\title{New Axioms for Dependence Measure and Powerful Tests}
\runtitle{Dependence Axioms \& Powerful Tests}

\begin{aug}
% use \particle for den|der|de|van|von (only lc!)
% [add1]{\fnms{}~\snm{}\ead[label=e?]{}}
%
%% e-mail is mandatory for each author
%
%%% initials in fnms (if any) with spaces
%
\author[add1]{\fnms{H. D.}~\snm{Vinod}\ead[label=e1]{vinod@fordham.edu}}
%%%%%%%%%%%%%%%%%%%%%%%%%%%%%%%%%%%%%%%%%%%%%%
%% Addresses                                %%
%%%%%%%%%%%%%%%%%%%%%%%%%%%%%%%%%%%%%%%%%%%%%%
\address[add1]{%
\orgdiv{Economics},
\orgname{Fordham University, 441 East Bronx Rd, New York 10458}}

\end{aug}

%% Put support info here.  Reminder: do not thank the handling coeditor anonymously or by name
%%%%%%We thank Fordham University.
%%%\end{funding}
%
\begin{abstract}
Statistical measure(s) of dependence (MOD)
between variables are essential for most empirical
work. We show that Renyi's postulates from the 1950s are 
responsible for serious MOD limitations.
(i) They rule out examples when
one of the variables is deterministic (like time or age),
(ii) They are always positive, implying no one-tailed significance tests. (iii) They disallow ubiquitous asymmetric MOD.
Many MOD exist in the literature, including those from 2022 and 2025, share these limitations because they fail
to satisfy our three new axioms. We also describe a new implementation of
a powerful one-sided test for the null of zero Pearson correlation with
Taraldsen's exact sampling distribution and provide
a new table for practitioners. We include a published
example where Taraldsen's test makes a practical difference.
The code to implement all our proposals is in R packages.
\end{abstract}


\end{frontmatter}
%%%%%%%%%%%%%%%%%%%%%%%%%%%%%%%%%%%%%%%%%%%%%%%%%%%%%%%%%%%%%%%%%%%%%%%%%
%%%% Main text entry area:
%%%%%%%%%%%%%%%%%%%%%%%%%%%%%%%%%%%%%%%%%%%%%%%%%%%%%%%%%%%%%%%%%%%%%%%%%

\section{Introduction}
\label{sec.intro}
A great deal of science focuses on understanding the
dependence between variables. Its quantification has a
long history, starting with the Galton-Pearson correlation
coefficient $r_{ij}$ from the 1890s and its cousins, including Spearman's $\rho$, Kendall's $\tau$, and Hoeffding's $D$.
\citet{Renyi59} argued that a measure of dependence
(MOD) should satisfy formal postulates irrespective of
specific contexts and applications.
Many measures, including recent \citet{Borgonovo25jrssb} 
(or ``BO25'') and the Hellinger measure by
\citet{GPierre22}(or ``GM22''), treat
many of Renyi's
postulates, especially symmetry,
as if they are sacrosanct. 
\subsection{List of Renyi's Postulates}
\begin{itemize}
	\item []
	P1) (Existence) Let $X_i$ and $X_j$ be two random variables on a probability space triplet where none is constant with
	probability 1. 
	\item [] 
	P2) (Symmetry) $MOD(X_i, \,X_j) = MOD(X_j, \,X_i)$. 
	
	\item [] 
	P3) (Positivity) $0\le MOD(X_i,\, X_j) \le 1$.
	
	\item [] 
	P4) $MOD(X_i,\, X_j)=0$ if and only if they are independent.
	
	\item [] 
	P5) $MOD(X_i,\, X_j)=1$ if and only if their
	dependence is strict in the sense that either
	$X_i$ or $X_j$ can be replaced by Borel measurable
	functions, $f(X_i)$, $g(X_i)$, $f(X_j)$, $g(X_j)$.
	
	\item [] 
	P6) If the Borel-measurable functions $f(x)$ and $g(x)$ map the real axis
	in a one-to-one way onto itself, then
	$MOD(f(X_i),\, g(X_j))=MOD(X_i,\, X_j)$.
	\item [] 
	P7) If the joint distribution of $X_i$ and $X_j$ is normal, then $MOD(X_i,\, X_j)$ measure equals the absolute
	value of the Pearson correlation coefficient $|r(i,j)|$.
\end{itemize}

We regard Renyi's postulates as
unsuitable for measuring many real-world dependence situations in 
natural or social sciences, medicine, and engineering.
Accordingly, our first task is to propose a revised set
of postulates suitable for a more inclusive (general) 
MOD.

Consider the dependence of a baby's 
weight on his age in weeks.
Renyi's existence
postulate (P1) excludes a study of such dependence
because one of the two variables,
age, is a non-random
sequence of deterministic numbers. 

\subsection{Some Definitions}
Our proposed MOD represents two
conditional measures that lack
reciprocity, since
$MOD(X_i|X_j) \not= MOD(X_j|X_i)$.  

\begin{definition}
	{\bf Max dependence}: We define
	\begin{equation}
		\label{eq.noconditioning}
		MaxMOD(X_i, X_j) = \max \{ |MOD(X_i|X_j)|, |MOD(X_j|X_i)| \},
	\end{equation}
	equal to the larger of the two conditional magnitudes. 
\end{definition}
BO25's measure of association is maximal if and only if we have 
a deterministic (noiseless) dependence. Measures of
association are 
different from our MOD.
\begin{definition}
	{\bf Small dependence}: We define
	$ |MaxMOD(X_i,\, X_j)|=\epsilon >0$  where the size of $\epsilon$ equals 
	a  ``small'' value (= 0.01, say) depending on the units and
	sampling variation. 
\end{definition}
BO25 recognize zero association if and only if random variables are independent. An 
advantage of BO25 is its derivation of rigorous asymptotic properties.
\begin{definition}
	{\bf Positive 
		dependence} of $X_i$ on $X_j$ as 
	$ MOD(X_i,\,| X_j) >\epsilon$, and similarly, 
	positive 
	dependence of $X_j$ on $X_i$ requires
	$ MOD(X_j \,| X_i) >\epsilon$.
\end{definition}

\begin{definition}
	{\bf Negative
		dependence} of $X_i$ on $X_j$ requires
	$ MOD(X_i,\,| X_j) < -\epsilon$. 
	The negative
	dependence of $X_j$ on $X_i$ requires
	$ MOD(X_j,\,| X_i) < -\epsilon$. 
\end{definition}

\subsection{New Axioms for Widely Applicable Dependence Measures}
Quantification of dependence  (by $MOD$) should follow certain 
general principles that are meaningful in
natural or social sciences, medicine, and engineering.
The following axioms are applicable in most contexts. This paper uses examples and logical arguments to show that failure to satisfy our
axioms generally leads
to an inferior $MOD$. Although Pearson correlations $R=\{r_{ij}\}$ satisfy these
axioms, they remain inferior to generalized correlations
$R^*=\{r^*_{i|j}\}$ defined later in Section \ref{sec.kernelReg}. 

%\begin{itemize}
%	\item [] 
%	{\bf A1)} 
\begin{axiom}[A1]
(Existence) $MOD$ is defined when numerical data on variables ($X_i ,\, X_j$)  exist.
\end{axiom}	
%	{\bf Remark A1)} 
\begin{remark}
%{\bf Remark A1)}
The data
variables need not have finite
	second moments and can be random or deterministic.
	Renyi's P1 disallows applications where 
	one of the variables is a time sequence or a computer-generated  random variable based on a deterministic seed. 
	Of course, dependence measures may not 
be meaningful when both variables are deterministic.
\end{remark}
%	\item [] 
%	{\bf A2}) 
\begin{axiom}[A2]	(Zero Dependence) When $MOD(X_i,\,| X_j)=0$, and
	$MOD(X_j,\,| X_i)=0$, we have zero dependence or
	independence.
\end{axiom}		
%	{\bf Remark A2)} 
\begin{remark}
%	{\bf Remark A2)}
	 Zero dependence
	is neither positive nor negative. Since
	exactly zero conditional dependence metrics
	are rare, let us use ``small dependence'' 
	defined above.
	There are 
	two conditions for ``practical'' independence,
	$|MOD(X_i,\,| X_j)|<\epsilon$, and
	$|MOD(X_j,\,| X_i)|<\epsilon$. See BO25 for
	asymptotically valid independence tests.
\end{remark}

%	\item [] 
%	{\bf A3)} 
\begin{axiom}[A3]
	(Restricted Range) The dependence measure must satisfy
	two range conditions:
	\begin{equation}
		\label{eq.bounds}
		-1\le  MOD(X_i\,| X_j) \le 1, \quad {and} \quad
		-1\le  MOD(X_j\,| X_i) \le 1.	
	\end{equation}
	$MOD=1$ is perfect positive dependence, and
	$MOD=-1$ is perfect negative (inverse) dependence.
\end{axiom}		
	%{\bf Remark A3)} 
\begin{remark}
%{\bf Remark A3)} 
The fixed range, $-1 \le MOD \le 1$,
	for all applications and in all contexts, making $MOD$ values
	comparable and providing crucial directional information in its sign. Known signs permit one-sided alternative
	null hypotheses leading to more powerful tests.
\end{remark}

%\end{itemize}

Since many continuous functions can be Borel
measurable $f(x)$ or $g(x)$,
Renyi's
postulates (P5) and (P6) that $MOD(X_i,\, X_j)$ should
remain invariant even after such transformations are not
applicable in more general contexts where the sign matters.
Note that a
change of sign is a Borel measurable transformation, and sign changes
reverse the direction of dependence. It is entirely
appropriate that our (A3) does not expect $MOD$ to remain
unchanged even after such direction reversals. 
Section \ref{sec.fishbirds} reports real-world
examples used by GM22, which better support our $MOD$.

\subsection{Four Examples of Asymmetric Dependence}
\label{sec.ex4}
This subsection contains our arguments
challenging Renyi's symmetry postulate (P2). We use examples
in nature or data where a correct metric for dependence cannot be symmetric. 
\begin{itemize}
	\item [i)] 
	A newborn baby boy depends on his mother for his survival, 
	but it is wrong to expect that his mother must exactly equally
	depend on the boy for her survival, as implied by (P2). 
	Consider data from a few geographical regions. Let $X_i$ be infant mortality, and $X_j$ be maternal deaths during childbirth. Insisting on exact equality $MOD(X_i |X_j) =  MOD(X_j |X_i)$ of 
	dependence metrics is absurd.
	\item [ii)]
	Meteorologists know that the average daily high of December temperatures
	in New York City is 44 degrees Fahrenheit. The number 44
	depends on New York's latitude (40.7).  Assume we have
	data on several city latitudes ($\ell$) and corresponding
	December temperatures ($\tau$). 
	Symmetric dependence by (P2) between temperature and latitude 
	implies $MOD(\ell |\tau) =  MOD(\tau |\ell)$.
	The latitude of a city does not depend on its
	temperature, $MOD(\ell |\tau)$ is near zero. P2 reciprocity amounts
	to the absurd requirement that a small number
	equal a large number.
	
	\item [iii)] For a third example, imagine a business person B owns several shops, not all doing equally well.
	B's 30\% earnings 
	depend on the hours a key employee works in one shop. Now, the symmetry  by (P2) means
	the absurd expectation that hours worked by the key employee
	(subject to labor laws)
	always depend on owner B's earnings, precizely 30\%.
	%even if some earnings are from another shop.
	\item [iv)]
	Our fourth example assumes $X_i$ as complete data, but its subset
	$X_j$ alone is available, and the rest of ($X_j\cap X_i$) is missing.  The available subset $X_j$ is a proxy that depends on $X_i$, but
	the complete set $X_i$ does not equally depend on its subset $X_j$.
\end{itemize}

\begin{proposition}
	We reject any dependence metric
	insisting on the exact equality $MOD(X_i|X_j) = MOD(X_j|X_i)$
	for all  $X_i$ and $X_j$. Of course, such equality can occur when  $X_i$ and $X_j$ are strictly linearly related in data.
\end{proposition}

\begin{proof}
	Use ``reductio ad absurdum.'' The four examples of Section
	\ref{sec.ex4} show the
	absurdity of denying Proposition 1. One can construct many more examples where variables need not be linearly
	related. Then, insisting on equality of $MOD(X_i|X_j)$ and $MOD(X_j|X_i)$ leads to untenable situations.
	
	Further examples include:
	$MOD(income |race) \ne MOD(race |income),\\
	MOD(stock \,price\, index |gdp) \ne MOD(gdp |stock\, price \, index),\\
	MOD(school\, test\, scores |income) \ne MOD(income |school \, test \, scores).$
\end{proof}
The 
symmetry postulate is neither necessary nor sufficient
for real-world dependence.  It is interesting that 
many authors have treated
Renyi's 
unrealistic postulate (P2) as
inviolable. 
The four examples above show why the symmetry postulate is 
absurd in many real-world contexts. These examples
comprise our main arguments for avoiding
the symmetry dogma. BO25's remark 2.17 needs to use
``separation measurements'' to obtain a symmetric
measure of association, distinguished from our MOD.

\subsection{Bivariate Linear Regression}
\label{sec.biv.linear}
Given data on $X_i$ and $X_j$, we can always
consider a bivariate linear regression model, $X_i=a+bX_j+\epsilon$. It is important to 
recognize that the statistical measure of
dependence of $X_i$ on $X_j$ is 
the coefficient of determination, $R^2_{i|j}$, not the
regression slope $b$. Similarly, one measures the
dependence of $X_j$ on $X_i$ by
$R^2_{j|i}$ of a flipped linear regression,
$X_j=a^\prime+b^\prime X_i+\epsilon^\prime$. 
The numerically exact, somewhat counterintuitive equality 
$R^2_{i|j}=R^2_{j|i}$ holds true despite distinct
slope coefficients ($b\ne b^\prime$).  The
equality of two flipped $R^2$ values is due to
the symmetry of covariances and linearity.
Our Section \ref{sec.ex4} and Proposition 1 avoid
the symmetry dogma.

Let us define
$r_{i|j}=\surd R^2_{i|j}$, where the sign of the square root is
equal to that of the covariance  $Cov(X_i, \,X_j)$.
We also define for the flipped regression
$\surd R^2_{j|i}= r_{j|i}$.

%\vspace{3mm}
%{\bf Corollary 1:}
\begin{corollary}
Assuming variables $X_i$ and $X_j$ are linearly related,
the Pearson correlation coefficient satisfies
$r_{ij}=r_{i|j}=r_{j|i}$.
\end{corollary}

%\vspace{3mm}
%{\bf Proposition 2:}
\begin{proposition}
	The Pearson correlation coefficient is a generally acceptable MOD.
\end{proposition}
\begin{proof} Since variables $(X_i,\, X_j)$ exist in flipped bivariate
	linear regressions, (A1) holds. Since 
	we interpret zero correlation as
	no dependence, (A2) holds. Since ($-1\le r_{ij}\le 1$)
	limits hold, A3 holds. The
	satisfaction of all axioms makes $r_{ij}$ generally acceptable.
\end{proof}

Over the last century, researchers have treated
the Pearson correlation coefficient
$r_{ij}$ as a first approximation to an acceptable measure of bivariate linear dependence. 
Our axioms A1 to A3 accept that practice. However, 
we draw the reader's attention to distinct
generalized correlations $r^*_{i|j}$ and  $r^*_{j|i}$, which relax the linearity assumption. See Section \ref{sec.kernelReg} for details.

Boyle's law states that a gas's pressure and volume are inversely proportional. Renyi's (P1) disallows a deterministic
sequence of volumes to the experimenter. A researcher in finance trying to model the
time-dependence of stock prices always has a deterministic time
sequence as one of the variables.
The existence of deterministic variables will also fail to satisfy
Renyi's (P7), because one deterministic variable will fail bivariate normality.
By contrast,
our existence axiom (A1) accepts deterministic variables and  
the generalized $r^*_{i|j}$ and $r^*_{j|i}$ include nonlinear dependence. 
Since a great many empirical studies have deterministic variables,
comprehensive $MOD$ should include them as our axioms do, and
Renyi's postulates do not.

The outline of the remaining paper is as follows. Section
\ref{sec.dogma} reviews the historical sources of the symmetry dogma.  Section \ref{sec.fifj} checks the
satisfaction of our axioms by
various dependence
measures in the literature. Many measures fail to satisfy the bounds in axiom A3.
Section \ref{sec.kernelReg} describes our preferred measures
from the matrix $R^*$ based on flipped kernel regressions.
Section \ref{sec.infer} discusses statistical inference for
correlation coefficients, including a new Table 1 for
one-sided (more powerful) inference using Taraldsen's distribution.
Section \ref{sec.examp} discusses examples explaining
the superiority of our axioms, and Section
\ref{sec.fin} contains final remarks.
\section{Sources of the Symmetry Dogma}
\label{sec.dogma}
Despite the examples in subsection \ref{sec.ex4}, why has
the symmetry dogma long persisted in statistics?
This subsection lists four plausible origins.
\begin{itemize}
	\item [] 
	(i) The definitional and numerical
	equality of covariances, $Cov(X_i,\, X_j) =Cov(X_j,\, X_i)$, may
	have been the initial reason for the symmetry result.
	\item [] 
	(ii) Recalling Section \ref{sec.biv.linear},
	the equality of two
	$R^2$ strengths supports the symmetry dogma. Considering
	the signed square roots of the two $R^2$ values, 
	the matrix $R=\{ r_{ji}\}$ also supports the dogma 
	under the harmless-looking
	linearity assumption. 
	Section \ref{sec.kernelReg} 
	in the sequel describes an asymmetric generalized correlation matrix $R^*=\{ r^*_{i|j}\}$ which avoids 
	the linearity and hence the dogma.
	
	\item [] 
	(iii) Since all distances satisfy symmetry, they have
	mathematical elegance and appeal. Such elegance may have been the
	reason for the wide acceptance of Renyi's symmetry postulate.  
	\item []
	(iv) The concept of statistical independence in probability
	theory is symmetric.
	It can be formulated in terms of the absence of any divergence
	between a joint density and a product of two marginal densities,
	\begin{equation}
		\label{eq.productOFmarginals}
		f(X_i \,X_j)= f(X_i)\, f(X_j).
	\end{equation}
	Since dependence is the opposite of independence,
	it is also tempting (but unhelpful) to impose symmetry on dependence.
	Symmetry is appropriate in Pearson's Chi-square
	test for statistical independence between row and
	column categories of a contingency table.
	
\end{itemize}

\section{New Axioms and Various Dependence Measures}
\label{sec.fifj}
Various
dependence measures in the literature are
designed to be helpful in certain contexts for certain types of data. The
next five subsections check whether the selected five
satisfy A1 to A3 to be called ``generally acceptable'' 
by avoiding the three limitations mentioned in the abstract.

\begin{proposition}
	%{\bf Proposition 3:}
	Dependence measures that fail to satisfy
	all three axioms A1 to A3 are 
	not generally acceptable. 
\end{proposition}
\subsection{Unacceptable Dependence Metric for Time Series}
\citet{Granger04} (or ``Gr04'') is an important paper on formal testing for
statistical independence, especially for time series data. They
cite a survey by \citet{Tjo96} on the topic.  The novelty in Gr04
is in using
nonparametric nonlinear kernel densities in testing the equality (\ref{eq.productOFmarginals}) in their test of independence.
Unfortunately, Gr04 authors
adhere to the symmetry dogma by 
insisting that their dependence metric 
should be a symmetric distance-type. Since always positive distances fail A3, Gr04's metric is unacceptable.

\subsection{Unacceptable Dependence Metrics Based on Entropy}
\label{sec.entropy}
Assume we have data
in the form of probabilities associated with specific values of
($X, \,Y$) variables obtained after the data are
sorted and split into a certain number
of class intervals. The corresponding frequencies relative to the
total frequency often define probability distributions
($f(x), \,f(y)$). Categorical data in contingency
tables also yield marginal and conditional probabilities
relevant for entropy computations.

Shannon defined information content in 1948 
as the amount of surprise in a
piece of information. His ``information'' is inversely proportional
to the probability of occurrence and applies to discrete and
continuous random variables with probabilities defined by
a probability distribution.

Intuitively, entropy is our ignorance or the extent of disorder in
a system. 
The entropy $H(Y)$ is defined by the mathematical expectation of
the Shannon information or $E(-\log \,f(y))$.

The conditional entropy of $Y$ given $X$, averaged over $X$, is
\begin{equation}
	\label{eq.hy|x}
	H(Y|X)=-E[ E[\log \,(f_{Y|X}(Y|X))|X]].
\end{equation}

The reduction in 
our ignorance
$H(Y)$ by knowing the proxy $X$ is $H(Y)-H(Y|X)$.
The entropy-based measure of dependence is
\begin{equation}
	\label{eq.entrop}
	D(X;Y)=\frac{H(Y)-H(Y|X)}{H(Y)},	
\end{equation}
or proportional reduction
in entropy of $Y$ by knowing $X$. The extreme values are
$D(X;Y)=0$ when $H(Y)=H(Y|X)$, and
$D(X;Y)=1$ when $H(Y|X)=0$. However, since (\ref{eq.entrop}) cannot
be negative, it fails to satisfy Axiom A3 and is unacceptable.

\vspace{4mm}
A related
measure of dependence in the entropy literature is
{\bf mutual information}, defined as
\begin{equation}
	\label{eq.mutu}
	I_{mu}(X, Y)=H(X)+H(Y)-H(X,Y).
\end{equation}
Since $I_{mu}(X,\, Y)=I_{mu}(Y, \,X)$, mutual information 
runs up against our examples in Section \ref{sec.ex4},
where symmetry is unacceptable. It, too, fails to satisfy 
axioms A2 and A3. Of course, entropy-based metrics are helpful
in specific contexts where $f(x)$ and $f(y)$ densities are
available, but remain generally unacceptable.

\subsection{Unacceptable Dependence Metric from Fisher Information}
\label{sec.Fisher}
Fisher information measures the expected amount of information 
given by a random variable $Y$ about a parameter $\theta$ of interest.
Under Gaussian assumptions, the Fisher information is inversely proportional
to the variance.
\citet{Reimherr13} use Fisher's information to define a measure of
dependence.
Consider estimating a model parameter $\theta$ using 
$X$ as a proxy for unavailable $Y$.  
$X$ is a subset of $Y$ with missing observations, as in the fourth
example of Section \ref{sec.ex4}. 
If the Fisher information for $\theta$ 
based on proxy $X$ is denoted by $\mathcal{I}_X(\theta)$, they
define a measure of dependence as:
\begin{equation}
	\label{eq.inforatio}
	D(X;Y)=\frac{\mathcal{I}_X(\theta)}{\mathcal{I}_Y(\theta)},	
\end{equation}
where $\mathcal{I}_X(\theta) \le \mathcal{I}_Y(\theta)$. 
Consider the special case where a proportion $p$
of the $Y$ data is missing in $X$ at entirely random  locations.
Then, the measure of dependence (\ref{eq.inforatio}) equals $0<p<1$. Since $D(X;Y)\ne D(Y;X)$,
this dependence measure
is asymmetric and useful when the focus is on
the estimation of a parameter $\theta$ from a
subset. Unfortunately, $D(X;Y)$
fails to admit negative values required by our axiom A3,
making it generally unacceptable.

\subsection{Unacceptable Dependence Metrics from Copulas}
\label{sec.copula}

Consider a two-dimensional joint (cumulative)
distribution function $F(X,\, Y)$ and marginal densities
$U=F_1(X)$ and $V=F_2(Y)$
obtained by probability integral transformations.
Sklar proved in 1959 that
a copula function $C(F_1,\, F_2)=F$ is unique if the components are
continuous. The copula function $C: [0, 1]^2\to [0,1]$ 
is subject to certain conditions, forcing it to be a
bivariate uniform distribution function. It is
extended to the multivariate case to
describe the dependence structure of the
dependence when row and column characteristics are
continuous variables rather than simple categories.

\citet{Dette13} (or ``DSS13'') define joint density as $F_{X, Y}$,
and the conditional density of $Y$ given $X$ as  $F_{Y|X=x}$.
They
use uniform random variables  $U$ and $V$ to construct a
copula $C$ as a
joint distribution function. The copula serves as their measure of dependence
based on the quality of regression-based prediction of $Y$ from $X$. Unfortunately, DSS13 ignore a
flipped prediction of $X$ from $Y$.

DSS13 assume
Lipschitz continuity, which implies that a copula is absolutely continuous
in each argument so that it can be recovered from any of its partial derivatives by
integration. The conditional distribution $F_{V|U=u}$ is related to
the corresponding copula $C(X, Y)$ by
$F_{V|U=u}(v)=\partial_1 C_{X,Y}(u, v)$.

A strictly symmetric measure of dependence proposed by DSS13 denoted 
with a subscript D as follows.
\begin{equation}
	\label{eq.rcopula}
	r_D(X,Y) =6 \int_0^1 \int_0^1 F_{V|U=u} (v)^2 dv du,
\end{equation}
where $r_D=0$ represents independence, and $r_D=1$ represents
almost sure functional dependence. DSS13 focus on $r_D$ filling
the intermediate range of the closed interval $[0,1]$ while ignoring the negative range $[-1, 0)$, failing to satisfy axiom A3. Hence, 
$r_D$ are generally unacceptable.
DSS13 rely on parametric copulas, making them
subject to identification problems,
as explained by \citet{Allen-jrfm22}. 
Remark 3.7 in \citet{Beare10} states that symmetric copulas imply
time reversibility, which is unrealistic for social science, economics, and financial data.

In closing this subsection on copulas, we
note examples where they can satisfy our axioms.
\citet{BOURI18} reject the symmetry dogma and note that their parametric
copula can capture tail dependence, which is essential in a study
of financial markets. 
\citet{Allen-jrfm22} uses nonparametric
copula construction and asymmetric $R^*$, which are
detailed in Section \ref{sec.kernelReg},
satisfying our axioms.

\subsection{Unacceptable Hellinger Correlation $\eta$}
\label{sec.hell}
Now, we turn to the recent GM22 paper mentioned earlier,
which proposes Hellinger correlation
$\eta$ as a new symmetric measure of the strength of dependence.  Unfortunately, their $\eta \notin [0,1]$
fails to satisfy Renyi's (P3).
Hence, GM22 introduce a normalized version $\hat\eta \in [0,1]$ of Hellinger correlation. An advantage of $\hat\eta$ over
Pearson's $r_{ij}$ is that it incorporates some
nonlinearities.

Let $F_1$ and $F_2$ denote the known marginal distributions
of random variables $X_1$ and $X_2$, and let $F_{12}$ denote
their joint distribution. Now, GM22 authors ask readers
to imagine reconstructing
the joint distribution from the two marginals. The
definition of the strength of dependence by GM22 
is the
size of the ``missing link'' in reconstructing the joint
from marginals. This definition allows GM22 to claim that
symmetry is ``unquestionable.''

GM22 authors define the squared Hellinger distance $\mathcal{H}^2(X_1,\, X_2)$
as the missing link between $F_{12}$ and $F_1 F_2$. They approximate
a copula formulation of $\mathcal{H}^2$ using
the \citet{Bhatta43} affinity coefficient $\mathcal{B}$.
Let $C_{12}$ denote the copula of $(X_1,\, X_2)$, and $c_{12}$ 
denote its density.
The computation of $\hat\eta$ in the R package {\bf HellCor} uses
numerical integrals
$\mathcal{B}=\int\int \surd c_{12}$. Hellinger correlation
$\eta$ is
\begin{equation}
	\label{eq.eta}
	\eta=\frac{2}{\mathcal{B}^2}\{ \mathcal{B}^4+(4-3\mathcal{B}^4)^{1/2}-2  \}^{1/2}.
\end{equation} 
The Hellinger correlation is symmetric, $\eta(X_1,X_2)=\eta(X_2, X_1)$.

GM22 state that their Hellinger correlation $\eta$ needs to be
normalized to ensure that $\eta \in [0,1]$ because their
estimate of $\mathcal{B}$ can exceed unity. 
GM22 denote
the normalized version with a hat as $\hat \eta$ and claim an
easier interpretation of $\hat\eta$  on the ``familiar Pearson scale,''
though Pearson's $r_{ij}\in [-1, 1]$ scale admits negative values.
GM22 employ considerable ingenuity to achieve the positive range
[0, 1] described in their Section 5.3. They 
state on page 650 that their range normalization ``comes at the price of
a lower power when it comes to testing for independence.''
GM22 provide an R package {\bf HellCor} to compute $\hat\eta$ from data  as a  measure of dependence and 
test the null
hypothesis of independence of two variables.

\section{Details of Acceptable $R^*$ to Measure Dependence}
\label{sec.kernelReg}
This section describes the details of generalized correlations and
why we recommend $R^*$ for measuring bivariate dependence.
We noted earlier that covariances satisfy
symmetry, $Cov(X_i,\,X_j) =Cov(X_j,\,X_i)$, and  symmetric covariances suggest the overall direction
of the dependence between the two variables. For example,
$Cov(X_i,\,X_j)<0$ means that when $X_i$ is relatively up (larger), $X_j$ is down (smaller) in most cases.
Most of the symmetric measures of dependence discussed above fail to
provide this type of useful directional information except for
Pearson's correlation coefficients $r_{ij}$. Hence, 
$r_{ij}$ has retained its popularity as a valuable measure
of dependence for over a century despite assuming 
unrealistic linearity.

\citet{Zheng2012} introduce non-symmetric
generalized measures of correlation ($GMC\in [0,1]$), proving that
\begin{equation}
	\label{eq.gmcs}
	GMC(X_i|X_j)\ne GMC(X_j|X_i).
\end{equation}
Since GMCs fail to provide directional information in covariances
needed by practitioners,
\citet{VinodRao14} extends \citet{Zheng2012} to develop two distinct
generalized correlation coefficients
$-1 \le r^*(X_i|X_j) \le  1$ and
$-1 \le r^*(X_j|X_i) \le  1$ depending on the conditioning
variable. Computing net dependence
after removing the effect of additional variables $X_k$ 
as control or nuisance variables and generalized 
partial and canonical correlations from
\citet{Vinod15b} and
\citet{VCompEcon2021} are outside the bivariate scope of this paper.

A bivariate nonlinear nonparametric kernel regression of $X_i$ on
$X_j$ is $X_i=f(X_j) + error $. Assuming 
$n$ observations, the algorithm first estimates $n$ values
of the
conditional expectation function $E(X_i |X_j)$ as the fitted
values. 
The coefficient of determination of this regression,
$R^2(X_i|X_j)$, is simply the squared correlation coefficient between
observed and fitted values of $X_i$. The
corresponding $\surd R^2$
yields  $r^*(X_i|X_j)= r^*(i|j)$.
The flipped kernel
regression of $X_j$ on
$X_i$ similarly yields  $R^2(X_j|X_i)$
and its square root yields
$r^*(j\,|i)$, assuming both regressions exist.

Consider an artificial example where
$Z$ is unit Normal, $X_j$ is Student's t with three
degrees of freedom, and
independent of $Z$. Now, define
$X_i=Z X_j$ as a product of two independent
random variables whose unconditional expectation is zero, since $E(Z)=0$. Also, 
its conditional expectation is zero, $E(X_i |X_j)=0$.
However, $r^*(i\,| j)=-0.508$ and
$r^*(j\,| i)=-0.462$. The example shows that it is
difficult to guess the values of $R^*$ components from theoretically known conditional expectation values.

The matrix $R^*$ with elements $\{r^*(i\,| j)\}$ uses the standard
designation $i$ for rows and $j$ for columns.
Nonparametric nonlinear free-form regressions generally have
superior fits (larger $R^2$). Hence, 
the magnitude of
max($\{r^*(i\,| j)\},\{r^*(i\,| j)\}$) is
generally larger than the Pearson correlation coefficient
$r(i\,j)$.  Note that
$r^*_{i|j}\ne r^*_{j|i}$ implies that
the $R^*$ matrix is asymmetric.

\begin{proposition}
	%\vspace{3mm}
	%{\bf Proposition 4:}
	Generalized correlation coefficients 
	($r^*_{i|j} \ne r^*_{j|i}$) are acceptable dependence
	measures.
\end{proposition}
%\vspace{3mm}
\begin{proof}
	Since $(X_i, \,X_j)$ data exist, (A1) holds. When both
	$r^*_{i|j}=0$ and $r^*_{j|i}=0$ are true, there
	is zero dependence by (A2). Similar to Pearson correlation
	coefficients, we have
	$-1\le r^*_{i|j}\le 1$ and $-1\le r^*_{j|i}\le 1$, hence
	the range constraint of (A3) is satisfied.
\end{proof}

\vspace{3mm}
The R package {\bf generalCorr} 
%by \citet{generalCorr} 
uses kernel regressions to
overcome the linearity of $r_{ij}$ from the {\bf np} package, \citet{nppackage},
which can handle kernel regressions
among both continuous and discrete variables.

A special case of
(\ref{eq.noconditioning}) in the present context is an
appropriately signed 
larger of the two generalized correlation magnitudes or
\begin{equation}
	\label{eq.MODeas}
	MOD(X_i, X_j)= MOD(i,j) = sgn* max(|r^*(i|j)|,|r^*(j|i)|), 
\end{equation}
where $sgn$ is the sign of the covariance between
the two variables.  
One can use the R package {\bf generalCorr} %\citet{generalCorr}
and the R function {\tt depMeas(,)} to estimate equation
(\ref{eq.MODeas}), which 
is not mentioned in \citet{VinodRao14}. We prefer
explicit conditioning stated as $r^*(X_i|X_j)$ and $r^*(X_j|X_i)$ for proper interpretation.

The {\bf generalCorr} package
functions for computing $R^*$ elements are {\tt rstar(x,y)} and {\tt gmcmtx0(mtx)}.
The latter converts a data matrix argument (mtx)  with $p$
columns into a $p\times p$ asymmetric matrix $R^*$ of generalized
correlation coefficients.  Regarding the direction of (causal) dependence,
the convention is that the variable named in the column is the ``cause''
or the right-hand regressor, and the variable named along the row
is the response.  Thus, the recommended dependence measures from $R^*$ are
easy to compute and interpret. See an application to forecasting 
the stock market index of fear (VIX) and causal path determination in \citet{AllenHooper18}.

\section{Statistical Inference for Correlation Measures}
\label{sec.infer}
%\section{Statistical Inference for $r^*$}

We recommend the signed generalized correlation
coefficients as elements of
the $R^*$ matrix as the best $MOD$.
Its advantages include the avoidance of
the potentially misleading symmetry dogma and a proper
measurement of arbitrary nonlinear dependence dictated by the data.
This section describes an additional advantage of $R^*$:
it
allows a more powerful 
(one-tailed) inference. We shall see 
in Section \ref{sec.fishbirds} an example of how higher power matters.

The sign of each element of the $R^*$ matrix equals the
sign of the covariance $Cov_{ij}=Cov(X_i,\, X_j)$. A two-tailed
test of significance is appropriate only when $Cov_{ij}$  is 0.
Otherwize, a one-tailed test is applicable.
Any one-tailed test provides greater power to detect an effect in one direction by not testing the effect in the other direction,
\citet{Kendallv2}, Sections 22.24 and 22.28.

Since the sample correlation coefficient $r_{ij}$
from a bivariate normal parent has a non-normal
distribution, Fisher
developed his famous $z$-transformation in the 1920s. He
proved 
that the following transformed statistic $r^T_{ij}$ is approximately normal
with a stable variance,
\begin{equation}
	\label{eq.fisherT}
	r^T_{ij}=(1/2) \quad log \frac{(1+r_{ij})}{(1-r_{ij})}
	\sim N(0, 1/n),
\end{equation}
provided $r_{ij}\ne 1$ and $r_{ij}\ne -1$.
Recent work has developed the exact distribution of a correlation
coefficient.  It is now possible to directly compute
a confidence interval for any hypothesized value $\rho$ of the
population correlation coefficient.

Let $r$ be the empirical correlation of a random sample of size $n$ from
a bivariate normal parent. Theorem 1 of \citet{Taraldsen} generalized Fisher's famous $z$-transformation, extended by C. R. Rao.
The exact sampling distribution with $v=(n-1)>1$ is
\begin{eqnarray}
	\label{eq.tara}
	f(\rho| r,v)&=\frac{v(v-1)\Gamma(v-1))}{\surd(2\pi)\Gamma(v+0.5)}(1-r^2)^{\frac{v-1}{2 }}\,
	(1-\rho^2)^{\frac{v-2}{2}} (1-r \rho)^{ \frac{1-2v}{2} } \\\nonumber
	& \quad F(\frac{3}{2};-0.5; v+0.5; \frac{1+r\rho}{2}),
\end{eqnarray}
where F(.;.;.;.) denotes the Gaussian hypergeometric function, available in the
R package {\bf hypergeom} by R.K.S Hankin. 
The R package {\bf practicalSigni}  contains
an R function {\tt qTarald()} for quantiles and {\tt  pvTarald()} for
p-values based on (\ref{eq.tara}) over a grid of $r$ values
used in constructing our Table \ref{tab.1} and Figures
\ref{fig.exactfr} and \ref{fig.exactfr2} below.

Assuming that the data come from a bivariate normal parent, 
the sampling distribution of
any
correlation coefficient is (\ref{eq.tara}). Hence,
the sampling distribution of unequal off-diagonal elements of
the matrix of generalized correlations $R^*$ also follows
(\ref{eq.tara}). When
we test the null hypothesis
$H_0: \rho= \rho_0$, the relevant sampling distribution 
is 
obtained by plugging $\rho= \rho_0$  in (\ref{eq.tara}), depicted in
Figure \ref{fig.exactfr},
for two selected sample sizes. Both distributions are centered at the zero null value $\rho_0=0$. Similarly,
plugging $\rho=0.5$ in (\ref{eq.tara})
is depicted in Figure \ref{fig.exactfr2}.  
A confidence interval is readily computed from
two quantiles of the sampling distributions.
If the hypothesized null value of the correlation coefficient is inside the confidence interval,
we say that the observed $r$ is statistically insignificant.

\begin{figure}[h]
	\caption{Taraldsen's exact sampling density of  a
		correlation coefficient under the
		null  of $\rho=0$, solid line $n$=50, dashed line $n$=15}
	\label{fig.exactfr}
	\includegraphics[height=2.5in,width=4.5in]{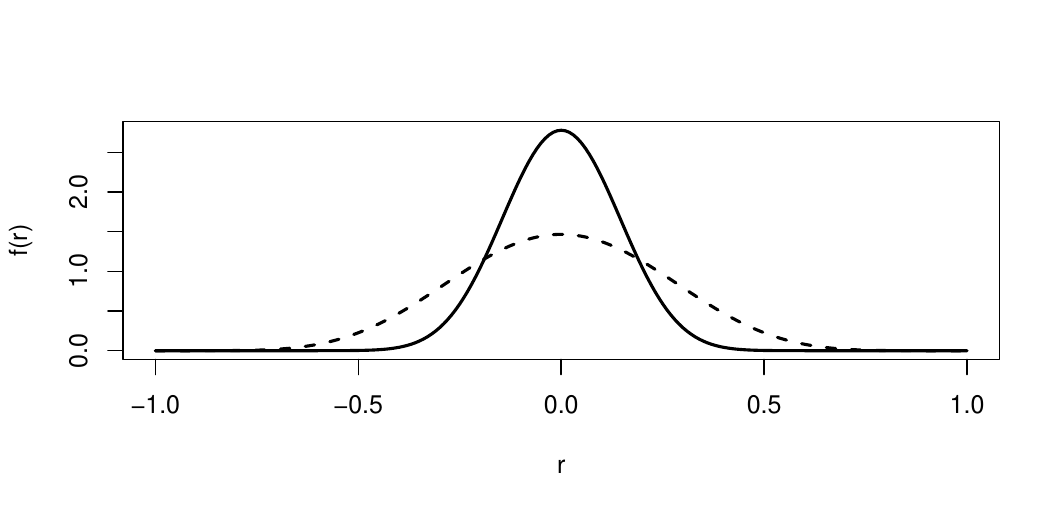}
\end{figure}

\begin{figure}[h]
	\caption{Taraldsen's exact sampling density of correlation coefficient under the
		null  of $\rho=0.5$, solid line $n$=50, dashed line $n$=15}
	\label{fig.exactfr2}
	\includegraphics[height=2.5in,width=4.5in]{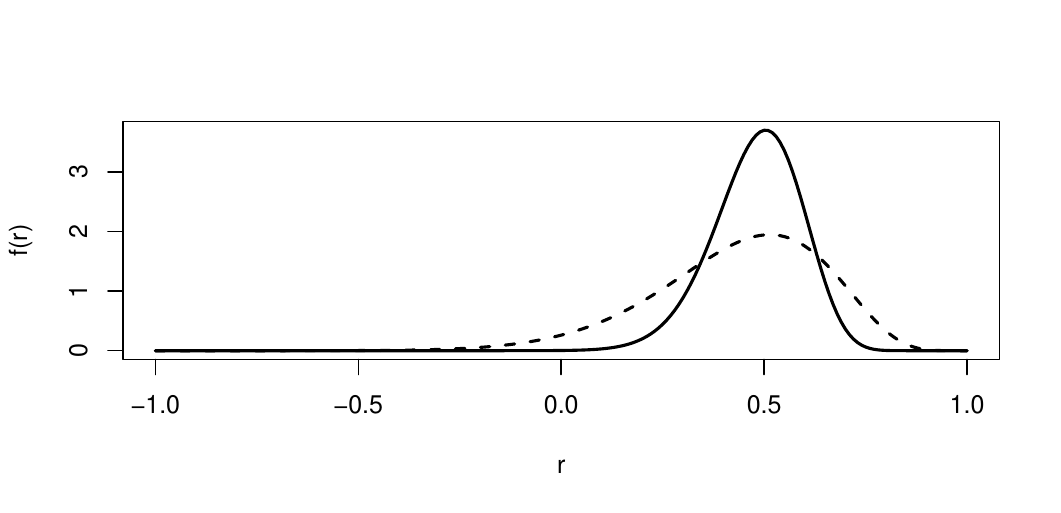}
\end{figure}

Taraldsen's exact densities depicted in
Figures \ref{fig.exactfr} and \ref{fig.exactfr2} 
depend on the sample size and 
the population correlation coefficient,  $-1 \le \rho \le 1$.
Given any hypothesized $\rho$ and sample size, a computer algorithm
readily computes the exact density, similar to Figures \ref{fig.exactfr} and \ref{fig.exactfr2}.  
We facilitate testing
the null hypothesis $\rho =0$ by creating a table of
a set of typical quantiles evaluated at specific 
cumulative probabilities and a corresponding selected set
of standard sample sizes.

Because of the complicated form of the density (\ref{eq.tara}),
it is not surprising that its (cumulative) distribution function $\int_{-1}^r f(\rho| r,v)$ by
analytical methods is not available in the literature. 
Hence, we compute cumulative probabilities by
numerical integration, defined as the rescaled area under the 
curve $f(r,v)$ for $\rho=0$. See Figure \ref{fig.exactfr}
for two sample sizes ($n$=50, 15) where $v =n-1$. 
The cumulative probability becomes a sum of 
rescaled areas of small-width rectangles
whose heights are determined by the curve tracing $f(r,v)$.
The accuracy of numerical approximation to the area is obviously
better if the number of rectangles is larger.

The R command {\tt r=seq(-1,1, by =0.001)} produces a sequence of $r\in [-1, 1]$, yielding 2001  rectangles
of width 0.001. Denote
the height of $f(r,v)$ by $H_f=H_{f(r,v)}$. 
Now, the area between any two 
$r\in [-1, 1]$ limits,  say $r_{Lo}$ and $r_{Up}$, is a 
summation of areas (height times width=0.001) of all rectangles. Now, the cumulative probabilities in the range are
\begin{equation}
	\label{eq.cumProb}
	\Sigma_{r_{Lo}}^{r_{Up}} H_{f} / \Sigma_{-1}^{1} H_f,
\end{equation}
where the common width cancels, and where the denominator
$ \Sigma_{-1}^{1} H_f$ converts the
rectangle areas into probabilities. We can
generally compute $\int f(\rho, r,v)$ for any $\rho \in [-1, 1]$.

Thus, we have a numerical approximation 
to the exact (cumulative) distribution function
under the bivariate normality of the parent,
$$F(\rho,r,v)=\int_{-1}^r f(\rho| r,v).$$ The transform
from $f(.)$ to $F(.)$ is called the probability integral
transform, and its inverse $F^{-1}(c|\rho, v)$ gives 
relevant correlation coefficients $r$ as quantiles
for specified cumulative probability $c$ as the argument. 
A computer algorithm finds such quantiles.

The exact $F^{-1}(c|\rho,v)$ allows the construction of confidence intervals
based on quantiles for each $\rho$ and sample size. 
For example, a 95\% two-tailed confidence
interval  uses the 2.5\% quantile $F^{-1}(c=0.025)$ as the lower limit
and 97.5\% quantile $F^{-1}(c=0.975)$ as the upper limit.
These limits depend on hypothesized $\rho$ and sample size.
Since $\rho=0$ is a common null hypothesis for correlation
coefficients,
let us provide a table of $F^{-1}(c)$ quantiles for
eleven sample sizes (listed in row names) and
eight cumulative probabilities 
listed in column titles of Table \ref{tab.1}.

The $p$-values in statistical inference
are defined as the probability of
observing the random variable
(correlation coefficient) as extreme or more extreme than
the observed value of the correlation coefficient $r$
for a given null
value $\rho=0$.  
Any one-tailed $p$-values based on
$f(\rho| r,v)$ of (\ref{eq.tara}) for arbitrary nonzero ``null'' values of $\rho$ can be similarly computed by numerical integration
defined as the area under the curve.
Use the
R function {\tt practicalSigni::pvTarald(.)}.

% latex table generated in R 4.2.1 by xtable 1.8-4 package
% Thu Nov 10 19:06:49 2022
\begin{table}[ht]
	\centering
	\caption{Critical values for higher-power 
		one-sided tests for Pearson's correlation r(i,j) when the null is
		$\rho=0$. We report quantiles
		evaluated at specified cumulative probabilities ($c$=.) using Taraldsen's exact sampling distribution for various sample sizes.}
	\label{tab.1}
	\begin{tabular}{rrrrrrrrr}
		\hline
		& $c$=0.01 & 0.025 & $c$=0.05 & $c$=0.1 & $c$=0.9 & $c$=0.95 & 0.975 & $c$=0.99 \\ 
		\hline
		$n$=5 & -0.83 & -0.75 & -0.67 & -0.55 & 0.55 & 0.67 & 0.75 & 0.83 \\ 
		$n$=10 & -0.66 & -0.58 & -0.50 & -0.40 & 0.40 & 0.50 & 0.58 & 0.66 \\ 
		$n$=15 & -0.56 & -0.48 & -0.41 & -0.33 & 0.33 & 0.41 & 0.48 & 0.56 \\ 
		$n$=20 & -0.49 & -0.42 & -0.36 & -0.28 & 0.28 & 0.36 & 0.42 & 0.49 \\ 
		$n$=25 & -0.44 & -0.38 & -0.32 & -0.26 & 0.26 & 0.32 & 0.38 & 0.44 \\ 
		$n$=30 & -0.41 & -0.35 & -0.30 & -0.23 & 0.23 & 0.30 & 0.35 & 0.41 \\ 
		$n$=40 & -0.36 & -0.30 & -0.26 & -0.20 & 0.20 & 0.26 & 0.30 & 0.36 \\ 
		$n$=70 & -0.27 & -0.23 & -0.20 & -0.15 & 0.15 & 0.20 & 0.23 & 0.27 \\ 
		$n$=90 & -0.24 & -0.20 & -0.17 & -0.14 & 0.14 & 0.17 & 0.20 & 0.24 \\ 
		$n$=100 & -0.23 & -0.20 & -0.16 & -0.13 & 0.13 & 0.16 & 0.20 & 0.23 \\ 
		$n$=150 & -0.19 & -0.16 & -0.13 & -0.10 & 0.10 & 0.13 & 0.16 & 0.19 \\ 
		\hline
	\end{tabular}
\end{table}

For the convenience of practitioners, we explain how
to use the cumulative probabilities
in Table \ref{tab.1} in the context of testing the null hypothesis $\rho=0$. A close look
at Table \ref{tab.1}
confirms that the distribution is symmetric around $\rho=0$, as in
Figure \ref{fig.exactfr}. Let us consider some examples.
If $n$=100, the critical value from Table 1 for a one-tailed 95\%
test is 0.16 (line $n$=100, column $c$=0.95). Let the observed
positive $r$ be 0.3. Since $r$ exceeds the critical value 
$(r>0.16$), we reject $\rho=0$.
If $n$=25, the critical value for a 5\% left tail in Table 1 is
$-0.32$. If the observed $r=-0.44$ is less than
the critical value $-0.32$, it falls in the left tail,
and we reject $\rho=0$ to conclude that it is significantly negative.

Table 1 can be used to construct two-tailed 95\% confidence intervals as follows.
If the sample size is 30, we see along the row $n$=30.  Now, column
$c$=0.025 gives $-0.35$ as the lower limit, and column $c$=0.975 gives
$0.35$ as the upper limit. In other words, for $n$=30, any correlation
coefficient smaller than 0.35 in absolute value is statistically
insignificant.

If the standard bivariate normality assumption is not believable, one can
use \citet{VinodJavier:2009}, the maximum entropy bootstrap (R package {\bf meboot}) designed for
dependent data.  A bootstrap application creates a large number 
$J=999$, say, versions
of data ($X_{i\ell}, \, X_{j\ell}$) for $\ell=1, \ldots J$. Each version
yields  $r^*(i|j;\ell), r^*(j|i;\ell)$ values. A large set of
$J$ replicates of these correlations gives  
a numerical approximation to the sampling distribution of these
correlations. Note that such a bootstrap sampling distribution is data-driven. Recall that
bivariate normality is needed 
for the construction of Table 1 based on (\ref{eq.tara}).

Sorting the replicated $r^*(i|j;\ell),\, r^*(j|i;\ell)$ values from the smallest to
the largest, one gets their ``order statistics'' denoted
upon inserting parentheses by replacing
$\ell$ by $(\ell)$. A
left-tail 95\% confidence interval for $r^*(i|j)$
leaves a 5\% probability mass
in the left tail. The interval is approximated by the order statistics as
$[r^*(i|j;(50)), 1]$. If the hypothesized $\rho=0$ is
inside the one-tailed interval, one fails to reject (accept) the null hypothesis 
$H_0: \rho=0$.

We conclude this section by noting that recommended measures of
dependence $MOD$ based on the $R^*$ matrix and their formal inference
are easy to implement. The tabulation of Taraldsen's exact sampling distribution
of correlation coefficients in Table 1 is new and deserves greater attention. 
The sampling distribution appears to be well-behaved, and limited interpolation and extrapolation of sample sizes and 
cumulative probabilities are possible.

We claim that Table 1, based on equations
(\ref{eq.tara}) and (\ref{eq.cumProb}), is an improvement over
textbook tables 
(or algorithms) for significance
testing of correlation coefficients based on Fisher's $z$-transform. 
We apply Table 1  and the
bootstrap inference discussed here to both older and
newer dependence measures.
The following section illustrates the superiority of
our axioms with published examples,
not handpicked for our purposes.

\section{Examples of Dependence Underestimation and Taraldsen tests}
\label{sec.examp}
Our first underestimation example deals with fuel economy 
using `mtcars' data in R software for 32 automobiles. We study the dependence between 
miles per gallon $mpg$, and horsepower $hp$.
\citet{VinodRao14} reports the Pearson correlation coefficient
$r(mpg,\, hp)=-0.78$, and two generalized correlation coefficients obtained by using
kernel regressions as
$r^*(mpg|hp)=-0.938$ and $r^*(hp|mpg)=-0.853$.
The  $MOD(mpg,\, hp)$ based on (\ref{eq.MODeas}) is
$-0.938$, showing the underestimation by the Pearson's correlation coefficient (=$-0.78$) due to assumed linearity.

Now, we illustrate using Talardsen's test.
Using an R function {\tt pvTarald(n=32, rho=0, obsr=-0.938)}
of the package {\bf practicalSigni} the one-tailed $p$-value is near zero, (= 1e-16). The fuel economy significantly depends on
horsepower. Practitioners
who do not wish to use R can
consult Table 1  column ``$c$=0.05'' for the five percent tests.
The row ``$n$=30'' for the sample size yields a
left-tail
critical value of $-0.30$. The observed correlation 
in the rejection region implies significant dependence.

Consider GM22's R package called
{\bf HellCor} for the same data. We find that
$\hat\eta=0.845>0$, giving no hint
that mpg and hp are negatively related. This is the
penalty for not obeying our axiom A3, which admits negative
$MOD$ when the variables are inversely related.

If we compare numerical magnitudes, we have
$\hat\eta=0.845$ larger than Pearson's $|r(mpg, hp)|=0.78$. 
This shows that $\hat\eta$ incorporates nonlinear dependence. 
However, $|\hat\eta|=0.845$ 
underestimates the magnitude $MOD$ ($=0.938$)
based on  (\ref{eq.MODeas}) and noted above. 
It seems plausible 
that the underestimation can be attributed
to the symmetry.
In addition to underestimation, Hellinger correlation's
symmetry postulate P2 means exact equality,
$MOD(mpg |hp) = MOD(hp |mpg)$, which is likely absurd
to auto engineers and car buffs.

\subsection{Further Real-Data Applications in GM22}
\label{sec.fishbirds}
GM22 claim superiority of Hellinger correlation over Pearson by
using two data sets. The underlying biological and demographic truth suggests significantly positive and negative correlations
for the two sets. A closer examination of the claim 
suggests important question marks.

Their first data set
refers to the population of
seabirds and coral reef fish residing around $n$ = 12 islands 
in the British Indian Ocean Territory of Chagos Archipelago.
Ecologists and other scientists cited by GM22 have determined
that fish and seabirds have an
ecologically symbiotic relationship. The seabirds create an
extra nutrient supply to help algae. Since fish primarily
feed on those algae, the two variables should have a significantly positive dependence. GM22 argue that the
underlying biology suggests a positive correlation, while the
statistical insignificance of the Pearson correlation
would violate the underlying biology.

GM22 begin with the low Pearson correlation $r(fish, \,seabirds)=0.374$
and a 95\% confidence interval  $[-0.2548,  0.7803]$
that contains a zero, suggesting
no significant dependence. 
The wide confidence interval, which
includes zero, is partly due to the small sample size ($n$=12).
The p-value using {\tt pvTarald(n=12, obsr=0.374)}
is 0.0935, which exceeds the benchmark of 0.05, confirming statistical insignificance. We agree with GM22's claim that Pearson's
correlation fails to support the biological truth.

Our Table 1 with the exact sampling distribution of correlations suggests that
when $n=10$ (more conservative than the correct $n$=12), the exact two-tailed 95\% confidence interval (leaving 2.5\% probability mass in 
both tails) 
also has
a wide range $[-0.58, 0.58]$, which includes zero. Assuming the direction is known, a one-tailed interval
with 5\% in the right tail ($n$=10) value is 0.50.  
It is significantly
positive (assuming a bivariate normal parent density)
only when the
observed correlation is larger than 0.50.

\begin{figure}[h]
	\caption{Marginal densities of fish and
		seabirds data are skewed, not Normal.}
	\label{fig.density}
	\includegraphics[height=2.5in,width=4.5in]{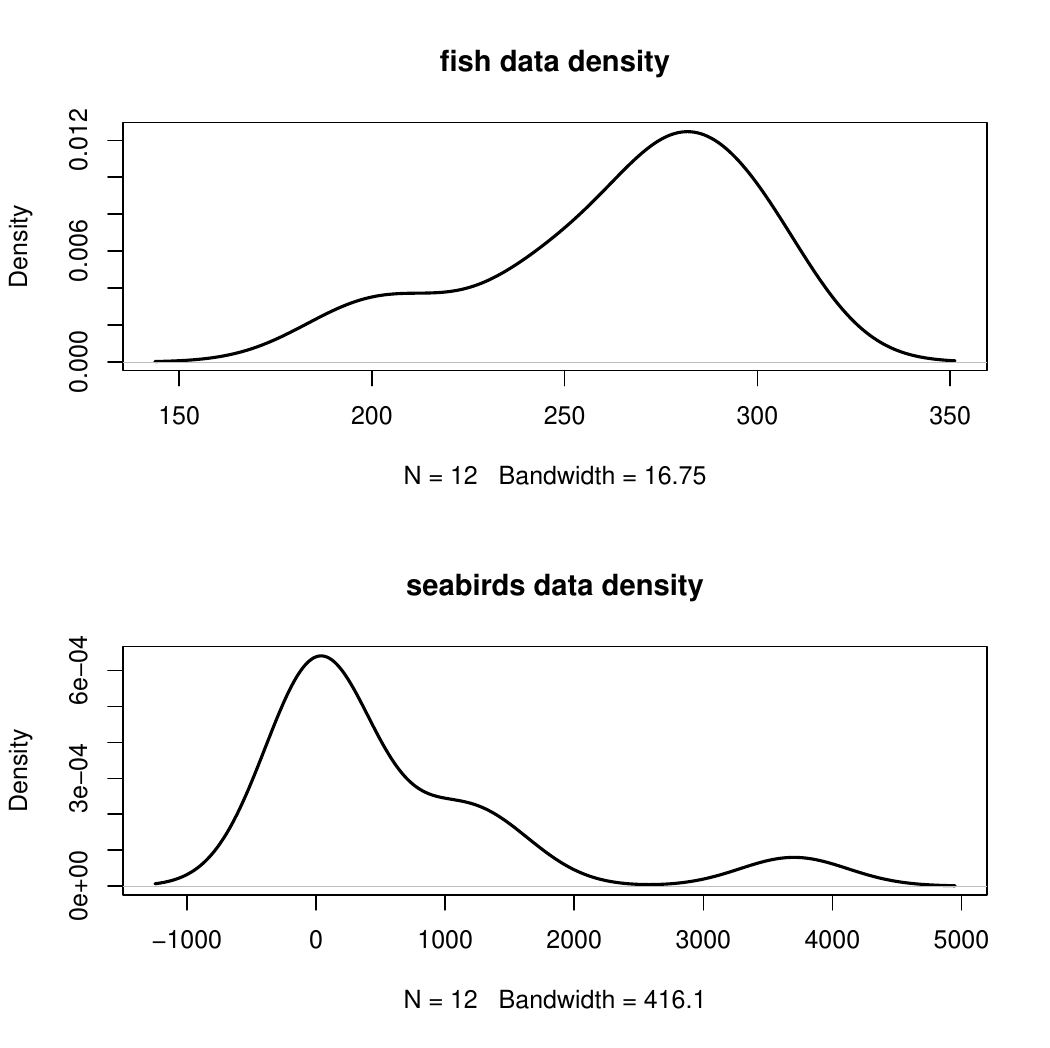}
\end{figure}

Using the population of
seabirds and coral reef fish residing around $n$ = 12 islands,  
GM22 report the estimate
$\hat \eta$(fish, seabirds)=0.744. 
Assuming a bivariate normal parent distribution
and using Taraldsen's exact density from Table 1,
$\hat\eta$(fish, seabirds)$=0.744 >0.50$, suggesting statistical significance.
The p-value using the R command {\tt pvTarald(n=12, obsr=0.744)}
is $0.0011 << 0.05$, indicating that the Hellinger correlation is highly significant.

Thus, the Hellinger correlation appears to support the biological truth, assuming a bivariate normal parent.
However, the GM22 authors report using the bootstrap
to relax the bivariate normality, which might
not hold for data with only $n=12$ observations.  
In light of the two marginal densities in Figure \ref{fig.density}, it is
unrealistic to
assume that the data come from a bivariate normal parent distribution.
Accordingly,
GM22 report a bootstrap p-value of
$0.045<0.05$ as their evidence.

Since GM22 bootstrap p-value of 0.045 is close to 0.05, 
it suggests  unintended p-hacking.
Let us redo their bootstrap.
When one runs their {\tt HellCor(.)} function with 
{\tt set.seed(99)} and default settings, the bootstrap p-value becomes $0.0513>0.05$, which
suggests insignificant $\hat\eta (fish, \, seabirds).$
Then, GM22's positive Hellinger correlation estimate of $\hat\eta$= 0.744 is not
statistically significant at the usual 95\% level in our bootstrap.
Thus, the Hellinger correlation fails to be convincingly
superior to Pearson's correlation $r$.
Both fail to confirm the biological truth because
both  $r$ and $\hat\eta$
may be insignificantly positive.

Let us compare $\hat \eta$ with our (\ref{eq.MODeas}) based on the
off-diagonal
elements of the generalized correlation matrix $R^*$ recommended here.
Our {\tt gmcmtx0 (cbind (fish, seabirds))} suggests the ``causal'' direction
(seabirds $\to$ fish) to be also positive, $r^*(fish| \, seabirds) =0.6687$. This causal direction from
$R^*$ agrees with
GM22's underlying biological truth mentioned above.
The p-value using {\tt pvTarald(..,obsr=0.6687)}
is $0.0044 < < 0.05$, confirming strong positive significance. 
We do not suspect p-hacking since the $p$-value 
(=0.0044) is not near 0.05. However, let us implement the
bootstrap as a robustness check.

\begin{figure}[h]
	\caption{Bootstrap density of generalized correlation coefficient
		r*(seabirds, fish).}
	\label{fig.bootdensity}
	\includegraphics[height=2.5in,width=4.5in]{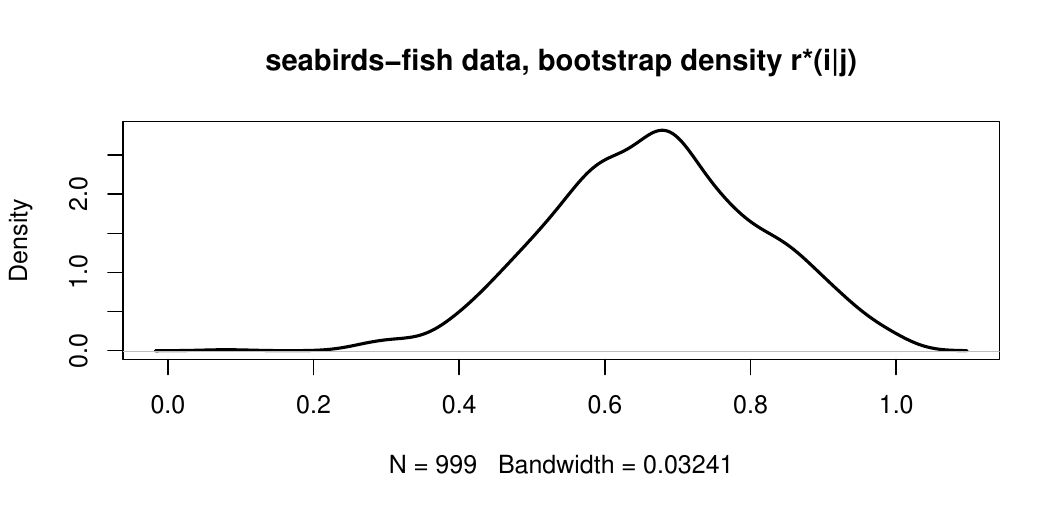}
\end{figure}

A 95\% bootstrap two-tailed confidence interval using the {\bf meboot} R package is
[0.3898, 0.9373]. A more powerful positive-tailed interval is [0.4394, 1], which also excludes zero. Even the lower limit of our {\bf meboot} confidence interval is not close to zero.
See Figure \ref{fig.bootdensity},
where almost the entire density has positive support.
Thus, the observed value is statistically significant and
positive, consistent with the 
biological truth, and establishes our axioms' superior performance in reaching the truth.
Also, our $MOD$ based on generalized correlation coefficients  $R^*$ satisfies A3 by revealing the
sign information hidden by the Hellinger correlation $\hat \eta$.

The second example in GM22 has
the number of births ($X_1$) and deaths ($X_2$) per year per 1000 individuals in $n$=229 countries in 2020.
A data scatterplot in their Figure 7 displays a C-shaped nonlinear 
relation. GM22 state (p. 651) that ``the strength of
this nonlinear structure of dependence is hardly captured
by Pearson's correlation.'' They explain that
$r(births,\, deaths)=-0.13$ is insignificant at level $\alpha = 0.05$. 
The  Hellinger correlation is $\hat\eta$ = 0.69 with a bootstrap 95\% all-positive confidence interval [0.474, 0.746], which correctly excludes
a zero, implying statistical significance. However,
the positive sign disagrees with the underlying demographic truth,
and may be confusing.

The statistical insignificance of the Pearson correlation
claimed by GM22 suffers from three
problems.
(a) GM22 use a less powerful two-tailed test of significance.
(b) GM22 rely on Fisher's $z$-transform for the
sampling distribution of the Pearson correlation coefficient. 
Their conclusion is reversed by
our more powerful one-tailed p-value using Taraldsen's exact sampling distribution, \citet{Taraldsen}.
Our R command {\tt pvTarald(n=229, obsr=-0.13)} based on
the {\bf practicalSigni} package yields a $p$-value of 0.0246.
Since $(0.0246<0.05)$,
Pearson
correlation (=$-0.13$) is
statistically significant. Thus, we have an example where
Taraldsen's density makes a practical difference, and the new result is
closer to the underlying demographic truth. (c)
A third problem with the $\hat \eta$ to 
measure dependence is that it hides the negative direction of
dependence, whereas the Pearson correlation does not.

\begin{figure}[h]
	\caption{Bootstrap density 
		of generalized correlation coefficient
		$r^*(death |birth)$.}
	\label{fig.birthdensity}
	\includegraphics[height=2.5in,width=4.5in]{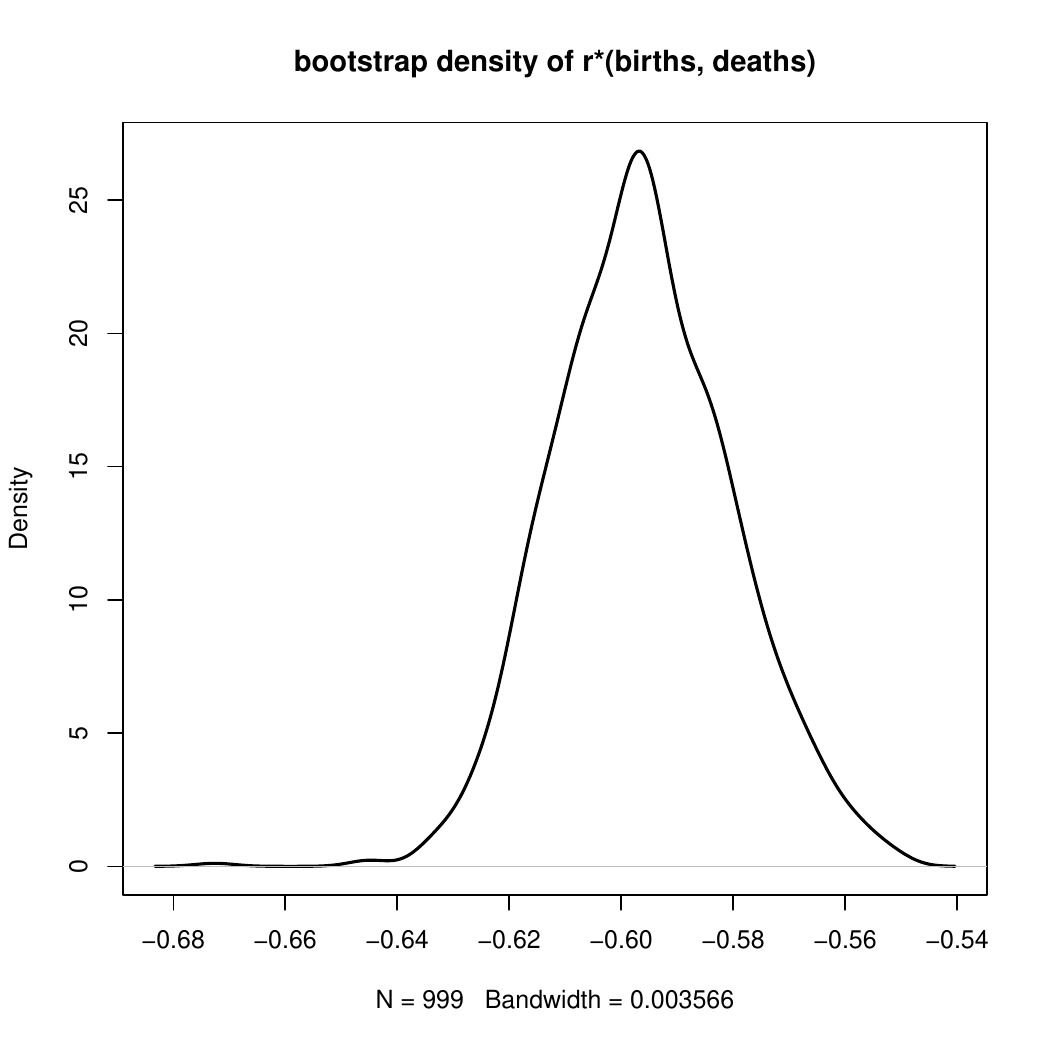}
\end{figure}

Let us estimate our $MOD$ of (\ref{eq.MODeas}) using the
data for GM22's second
example. The R command {\tt gmcmtx0( cbind(birth, death))} estimates that
$r^*(death|birth)$ is $=-0.6083$. A one-tailed 95\% confidence
interval using the maximum entropy bootstrap (R package {\bf meboot}) 
is [-1, -0.5693]. 
A less powerful
two-tailed interval [$-0.6251, -0.5641$]  is also entirely negative.
Since
our random intervals exclude zero, our $MOD$ is significantly 
negative. The $p$-value is near zero in Figure \ref{fig.birthdensity}
since almost the entire density has negative support.
A larger birth rate significantly leads to a lower death rate in
229 countries in 2020.

In summary, the two examples used by GM22 to sell their Hellinger correlation
have a discernible advantage over Pearson's $r_{ij}$ but not
over our $MOD$ based on generalized correlations $R^*$
satisfying our axioms. The examples confirm
five shortcomings of ``normalized''
Hellinger's correlation $\hat\eta$ over our
$MOD$ based on $R^*$. We have shown that
$\hat\eta$ can  (a) mislead, (b) underestimate, (c) hide
directional information, (d) disallow one-tailed
powerful tests, and (e) disallow deterministic
variables. Thus, satisfying our axioms is better than
satisfying Renyi's postulates.

\section{Final Remarks} 
\label{sec.fin}
Econometricians and other scientists are interested
in measure(s) of dependence (MOD) between variables. We show that
using Renyi’s seven 
postulates from the 1950s to define MOD implies three issues. (i) Admission of
deterministic variables.
(ii) Admission of one-sided tests of significance for greater power.
(iii) Avoidance of absurd implications of symmetric MOD. For example, insisting
that the dependence of the city temperature on its latitude should exactly
equal the (near-zero) dependence of the city latitude on its temperature. Sections \ref{sec.ex4} and \ref{sec.examp} have more examples.
It is hard to believe that
many researchers continue to ignore the absurdity.

Our propositions prove that
elements of the Pearson correlation matrix $R$ and
its generalized
version $R^*$ satisfy
our three axioms, whereas many others do not.
The R package {\bf generalCorr} and its 
vignettes make it easy to
compute and interpret $R^*$.
The
off-diagonal elements
of the asymmetric $R^*$ matrix  quantify dependence of
the row variable $X_i$ conditioned on
the column variable $X_j$, based on 
nonlinear and nonparametric relations among them.

Another novelty of this paper is implementing 
Taraldsen's alternative to Fisher's
$z$-transformation 
for the exact sampling distribution of correlation coefficients, plotted
in Figures \ref{fig.exactfr} and \ref{fig.exactfr2}. 
The R package {\bf practicalSigni}  contains
an R function {\tt qTarald()} for quantiles.
Our new
Table 1 provides new critical values for powerful 
one-sided tests for Pearson's $r(i,j)$
and generalized $r^*(i ,|j)$ when the null is a zero
population value ($\rho=0$), under bivariate normality.
Figures \ref{fig.bootdensity}
and \ref{fig.birthdensity} plot 
bootstrap sampling distributions for two examples
when the bivariate normality assumption is relaxed.

Interestingly,
our more exact inference matters for GM22's second
example, where the Pearson correlation $r(birth,\, death)$ is insignificant
by traditional methods but  
significantly negative using Taraldsen's density. Hence,
the complicated Hellinger correlation inference is unnecessary 
to achieve correct significance. Thus, both
handpicked examples designed to show the superiority of GM22's $\hat\eta$ over $r_{ij}$ also
show the merit of our proposal based on $R^*$ over $\hat\eta$. Our new axioms are an objective way of judging
statistical measures of dependence.

Almost every issue of every quantitative journal refers to correlation
coefficients at least once, underlining its importance in measuring dependence. Our Table 1 is relevant in a great many
situations for testing the significance of correlations and for
our $MOD$ based on $R^*$, satisfying
our three axioms. We hope these methods implemented in R
packages receive further attention,
usage, and development.

%%%%%%%%%%%%%%%%%%%%%%%%%%%%%%%%%%%%%%%%%%%%%%
%% Single Appendix:            %%
%%%%%%%%%%%%%%%%%%%%%%%%%%%%%%%%%%%%%%%%%%%%%%
%\begin{appendix}
%\section*{???} %% if no title is needed, leave empty \section*{}.
%\end{appendix}
%%%%%%%%%%%%%%%%%%%%%%%%%%%%%%%%%%%%%%%%%%%%%%
%% Multiple Appendixes:        %%
%%%%%%%%%%%%%%%%%%%%%%%%%%%%%%%%%%%%%%%%%%%%%%
%\begin{appendix}
%\section{???}
%
%\section{???}
%\end{appendix}

%%%%%%%%%%%%%%%%%%%%%%%%%%%%%%%%%%%%%%%%%%%%%%
%% Bibliography:                            %%
%%%%%%%%%%%%%%%%%%%%%%%%%%%%%%%%%%%%%%%%%%%%%%
%% IMPORTANT: References in the bibliography should be complete, 
%% including the first and last names, and date of publication.

%% If your bibliography is in bibtex format, uncomment commands:
%\bibliographystyle{ecta-fullname} % Style BST file
%\bibliography{ref12.bib}  % Bibliography file (usually '*.bib')

\begin{thebibliography}{20}
	\providecommand{\natexlab}[1]{#1}
	\providecommand{\url}[1]{\texttt{#1}}
	\expandafter\ifx\csname urlstyle\endcsname\relax
	\providecommand{\doi}[1]{doi: #1}\else
	\providecommand{\doi}{doi: \begingroup \urlstyle{rm}\Url}\fi
	
	\bibitem[Allen(2022)]{Allen-jrfm22}
	D.~E. Allen.
	\newblock Cryptocurrencies, diversification and the covid-19 pandemic.
	\newblock \emph{Journal of Risk and Financial Management}, 15\penalty0 (3),
	2022.
	\newblock ISSN 1911-8074.
	\newblock \doi{10.3390/jrfm15030103}.
	\newblock URL \url{https://www.mdpi.com/1911-8074/15/3/103}.
	
	\bibitem[Allen and Hooper(2018)]{AllenHooper18}
	D.~E. Allen and V.~Hooper.
	\newblock Generalized correlation measures of causality and forecasts of the
	{VIX} using non-linear models.
	\newblock \emph{Sustainability}, 10\penalty0 ((8): 2695):\penalty0 1--15, 2018.
	\newblock \doi{10.3390/su10082695}.
	\newblock URL \url{https://www.mdpi.com/2071-1050/10/8/2695}.
	
	\bibitem[Beare(2010)]{Beare10}
	B.~K. Beare.
	\newblock Copulas and temporal dependence.
	\newblock \emph{Econometrica}, 78(1):\penalty0 395--410, 2010.
	
	\bibitem[Bhattacharyya(1943)]{Bhatta43}
	A.~Bhattacharyya.
	\newblock On a measure of divergence between two statistical populations
	defined by their probability distributions.
	\newblock \emph{Bulletin of the Calcutta Mathematical Society}, 35:\penalty0
	99--109, 1943.
	
	\bibitem[Borgonovo et~al.(2025)Borgonovo, Figalli, Ghosal, Plischke, and
	Savaré]{Borgonovo25jrssb}
	E.~Borgonovo, A.~Figalli, P.~Ghosal, E.~Plischke, and G.~Savaré.
	\newblock Convexity and measures of statistical association.
	\newblock \emph{Journal of the Royal Statistical Society Series B}, page
	qkaf018, 04 2025.
	\newblock \doi{10.1093/jrsssb/qkaf018}.
	\newblock URL \url{https://doi.org/10.1093/jrsssb/qkaf018}.
	
	\bibitem[Bouri et~al.(2020)Bouri, Shahzad, Roubaud, Kristoufek, and
	Lucey]{BOURI18}
	E.~Bouri, S.~J.~H. Shahzad, D.~Roubaud, L.~Kristoufek, and B.~Lucey.
	\newblock Bitcoin, gold, and commodities as safe havens for stocks: New insight
	through wavelet analysis.
	\newblock \emph{The Quarterly Review of Economics and Finance}, 77:\penalty0
	156--164, 2020.
	\newblock ISSN 1062-9769.
	\newblock \doi{https://doi.org/10.1016/j.qref.2020.03.004}.
	\newblock URL
	\url{https://www.sciencedirect.com/science/article/pii/S1062976920300326}.
	
	\bibitem[Dette et~al.(2013)Dette, Siburg, and Stoimenov]{Dette13}
	H.~Dette, K.~F. Siburg, and P.~A. Stoimenov.
	\newblock A copula-based nonparametric measure of regression dependence.
	\newblock \emph{Scandinavian Journal of Statistics}, 40:\penalty0 21--41, 2013.
	
	\bibitem[Geenens and de~Micheaux(2022)]{GPierre22}
	G.~Geenens and P.~L. de~Micheaux.
	\newblock The hellinger correlation.
	\newblock \emph{Journal of the American Statistical Association}, 117
	(538):\penalty0 639--653, 2022.
	\newblock URL \url{10.1080/01621459.2020.1791132}.
	
	\bibitem[Granger et~al.(2004)Granger, Maasoumi, and Racine]{Granger04}
	C.~W.~J. Granger, E.~Maasoumi, and J.~Racine.
	\newblock A dependence metric for possibly nonlinear processes.
	\newblock \emph{Journal of Time Series Analysis}, 25:\penalty0 649--669, 2004.
	
	\bibitem[Hayfield and Racine(2008)]{nppackage}
	T.~Hayfield and J.~S. Racine.
	\newblock Nonparametric econometrics: The np package.
	\newblock \emph{Journal of Statistical Software}, 27\penalty0 (5):\penalty0
	1--32, 2008.
	\newblock URL \url{http://www.jstatsoft.org/v27/i05/}.
	
	\bibitem[Kendall and Stuart(1977)]{Kendallv2}
	M.~Kendall and A.~Stuart.
	\newblock \emph{The Advanced Theory of Statistics}, volume~2.
	\newblock New York: Macmillan Publishing Co., 4 edition, 1977.
	
	\bibitem[Reimherr and Nicolae(2013)]{Reimherr13}
	M.~Reimherr and D.~L. Nicolae.
	\newblock {On Quantifying Dependence: A Framework for Developing Interpretable
		Measures}.
	\newblock \emph{Statistical Science}, 28\penalty0 (1):\penalty0 116--130, 2013.
	\newblock URL \url{https://doi.org/10.1214/12-STS405}.
	
	\bibitem[Renyi(1959)]{Renyi59}
	A.~Renyi.
	\newblock On measures of dependence.
	\newblock \emph{Acta Mathematica Academiae Scientiarum Hungarica}, 10:\penalty0
	441--4510, 1959.
	
	\bibitem[Taraldsen(2021)]{Taraldsen}
	G.~Taraldsen.
	\newblock Confidence in correlation.
	\newblock \emph{arxiv}, pages 1--7, 2021.
	\newblock URL \url{10.13140/RG.2.2.23673.49769}.
	
	\bibitem[Tjostheim(1996)]{Tjo96}
	D.~Tjostheim.
	\newblock Measures and tests of independence: a survey.
	\newblock \emph{Statistics}, 28:\penalty0 249--284, 1996.
	\newblock URL \url{https://arxiv.org/pdf/1809.10455.pdf}.
	
	\bibitem[Vinod(2014)]{VinodRao14}
	H.~D. Vinod.
	\newblock Matrix algebra topics in statistics and economics using {R}.
	\newblock In M.~B. Rao and C.~R. Rao, editors, \emph{Handbook of Statistics:
		Computational Statistics with {R}}, volume~34, chapter~4, pages 143--176.
	North-Holland, Elsevier Science, New York, 2014.
	
	\bibitem[Vinod(2017)]{Vinod15b}
	H.~D. Vinod.
	\newblock Generalized correlation and kernel causality with applications in
	development economics.
	\newblock \emph{Communications in Statistics - Simulation and Computation},
	46\penalty0 (6):\penalty0 4513--4534, 2017.
	\newblock URL \url{https://doi.org/10.1080/03610918.2015.1122048}.
	\newblock Available online: 29 Dec 2015.
	
	\bibitem[Vinod(2021)]{VCompEcon2021}
	H.~D. Vinod.
	\newblock Generalized, partial and canonical correlation coefficients.
	\newblock \emph{Computational Economics}, 59(1):\penalty0 1--28, 2021.
	
	\bibitem[Vinod and {L\'opez-de-Lacalle}(2009)]{VinodJavier:2009}
	H.~D. Vinod and J.~{L\'opez-de-Lacalle}.
	\newblock Maximum entropy bootstrap for time series: The {meboot} {R} package.
	\newblock \emph{Journal of Statistical Software}, 29\penalty0 (5):\penalty0
	1--19, 2009.
	\newblock URL \url{http://www.jstatsoft.org/v29/i05/}.
	
	\bibitem[Zheng et~al.(2012)Zheng, Shi, and Zhang]{Zheng2012}
	S.~Zheng, N.-Z. Shi, and Z.~Zhang.
	\newblock Generalized measures of correlation for asymmetry, nonlinearity, and
	beyond.
	\newblock \emph{Journal of the American Statistical Association}, 107\penalty0
	(499):\penalty0 1239--1252, September 2012.
	
\end{thebibliography}

%% Or include bibliography directly:

% \begin{thebibliography}{}

% \end{thebibliography}

\end{document}